\documentstyle[prc,epsfig,aps]{revtex}

\begin{document}
\draft
\title{Faddeev treatment of 
long-range correlations and the one-hole spectral function of ${}^{16}{\rm O}$ }
\author{C. Barbieri${}^1$ and W.~H.~Dickhoff${}^{1,2}$}
  \address
     {${}^1$Department of Physics, Washington University,
	 St.Louis, Missouri 63130, USA \\  }
  \address
     {${}^2$Laboratory of Theoretical Physics, Ghent University,
	   Proeftuinstraat 86, B-9000 Gent, Belgium \\~\\  }

\date{\today}
\maketitle

\begin{abstract}
 The Faddeev technique is employed to study the influence of both
particle-particle and particle-hole phonons on the  one-hole spectral
function of ${}^{16}{\rm O}$.
 Collective excitations are accounted for at a random phase approximation
level and subsequently summed to all orders by the Faddeev equations to
obtain the nucleon self-energy.
An iterative procedure is applied to investigate the effects
of the self-consistent inclusion of the fragmentation in the determination
of the phonons and the corresponding self-energy.
The present results indicate that the characteristics of hole fragmentation
are related to the low-lying states of ${}^{16}{\rm O}$.
\end{abstract}

\section{Introduction}
\label{sec:1}
Correlations in the nucleus produce a substantial
reduction of the occupation probability of single-particle (sp) shells with
respect to the independent-particle model (IPM) prediction.
A substantial part of this depletion is due to the coupling to high-lying 
excitations reached by the strong short-range and tensor components of the
nucleon-nucleon interaction~\cite{bv91,dimu}.
At low energy, the presence of various collective modes may result in a
rearrangement of the sp strength distribution around the Fermi 
energy~\cite{rij92}.
Experimentally, these features can be observed in the $(e,e'p)$
reaction as a reduction of the absolute spectroscopic factors for the knockout
of a nucleon from a valence shell, as the strong fragmentation of the spectral
strength for more deeply bound sp states, and by the appearance of small
fragments associated with sp states that are empty in the IPM.
Studies of $(e,e'p)$ reactions have determined absolute spectroscopic
factors in many closed-shell nuclei~\cite{diep,sick,Lapik} demonstrating
that the removal probability for nucleons from these systems is reduced by
about 35\% with respect to the shell-model predictions.
 More recently, also $(e,e'NN)$ reactions, that involve the emission of two
nucleons from the target, have also become feasible~\cite{ond1,ond2}.
 The latter were motivated by the possibility to directly study the
high-momentum components, produced by short-range and tensor
correlations (SRTC), between pairs of nucleons at low energy. 
Such experiments are now able to disentangle the most relevant lowest
states of the residual $A-2$ nucleus~\cite{exp2} which are also
influenced by the presence of low-energy correlations.

The theoretical study of one- and two-hole spectral functions to understand the
results of the above reactions, usually requires substantial efforts in
computational many-body physics and does not always give a complete explanation
of the observed data.  In particular, the nucleus of ${}^{16}{\rm O}$ is
still not completely understood at the microscopic level and theoretical
calculations of its properties still fail in two ways.

 First, the experimental spectroscopic strength~\cite{leus}
for the knockout of a proton from both the $p_{1/2}$ and $p_{3/2}$ shells
corresponds to about $60\%$. 
The outgoing proton is described by a wave distorted by a complex
optical potential which describes the corresponding elastic proton 
scattering data.
This introduces an uncertainty in these spectroscopic 
factors since the $(e,e'p)$ reaction probes the interior of the nucleus,
where elastic proton scattering is less sensitive.
The inclusion of a relativistic description of the outgoing proton
in the analysis of the data can further change these results 
somewhat~\cite{udias}. An assignment of a 5 to 10\% error for the
absolute spectroscopic factors appears quite reasonable at this time.
Several studies have focused on the effect of short-range
correlations~\cite{md94,rad94,Fabr01} and computed spectroscopic factors
directly for ${}^{16}{\rm O}$. All these works yield a strength reduction
of about 10\% in general agreement with expectations based on nuclear-matter
calculations~\cite{bv91}.
Center-of-mass corrections are known to raise the spectroscopic factor
by about 7\%~\cite{diedef,O16jast}, resulting in a substantial disagreement
with data. 
The effects of long-range correlations (LRC) were studied in
Refs.~\cite{GeurtsO16,muetherO16}.
The inclusion of LRC in these works is limited to the Tamm-Dancoff (TDA)
approximation for
the intermediate two-particle$-$one-hole (2p1h) and two-hole$-$one-particle
(2h1p) states in the nucleon self-energy using the G-matrix as a residual
interaction. 
The reduction of the $p_{1/2}$ and $p_{3/2}$ spectroscopic factors due to these
LRC is about 15\%.
In the calculations of Ref.~\cite{GeurtsO16}, a combined treatment of LRC and 
SRTC was obtained.
The resulting $p_{1/2}$ and $p_{3/2}$ spectroscopic factors correspond to
77\% and 76\% of the sp strength, respectively, without inclusion of
center-of-mass corrections.
Short-range effects were included by employing the energy-dependence of the
Brueckner G-matrix in the corresponding self-energy contribution. 
The resulting disagreement with the experimental data appears to be
about 15-20\% but some allowance for the uncertainty of the extraction of the 
spectroscopic factors should be factored in. 
The results of Refs.~\cite{GeurtsO16,muetherO16} clearly show that 
a substantial improvement can be obtained by the inclusion of LRC.
Looking at the overall picture, a comparison between the quoted results
suggests that the quenching produced by SRTC appears to be well established.
At the same time, low-energy LRC are identified as an essential ingredient
that is needed for a complete understanding of the discrepancy with the
experimental data. 

The second issue regarding ${}^{16}{\rm O}$ concerns a satisfactory
understanding of the fragmentation of the sp strength at low energy.
The full one-hole spectral function for small missing energies
was computed in Ref.~\cite{GeurtsO16} and the results were 
subsequently used as a starting
point for the study of the $^{16}$O(e,e$'$pp) reaction~\cite{geurts2h,geueepp}. 
The high-momentum components of the two-hole overlap function caused by
short-range effects were also included by adding the proper defect functions
computed directly for the ${}^{16}{\rm O}$ nucleus~\cite{CALGM}.
These calculations led to a proper description of the experimental cross
section for two proton emission~\cite{ond1,ond2} to the ground state of
${}^{14}{\rm C}$.
 In spite of these successes, the one- and two-hole spectral functions still
miss some key features of the fragmentation and do not describe all the relevant
low-lying states of the residual nuclei ${}^{15}{\rm N}$ and ${}^{14}{\rm C}$.
In particular, the experiments show that the spectral strength for the
emission of a $p_{3/2}$ proton is fragmented in one big peak and a few 
smaller ones~\cite{leus}. The latter are found at a slightly
higher missing energy and carry a reasonable amount of strength: of the 60\%
observed mean-field strength, about 5\% is distributed in these small peaks
while the rest is in the main fragment.
This feature is not reproduced by the results of 
Refs.~\cite{GeurtsO16,muetherO16}, where
all the strength is concentrated in one single peak.
 Other experimentally observed hole fragments, like the $d_{5/2}$ and
$s_{1/2}$ found at about -17.4~MeV missing energy, are not reproduced as well.

  Analogously, the spectrum of ${}^{14}{\rm C}$ contains two low-lying
isovector $2^+$ levels, that can be reached by two-nucleon emission, but only
one of them is reproduced by the above theoretical calculations.
This missing state
represents the main discrepancy between the theoretical and experimental
$^{16}$O(e,e$'$pp) cross sections~\cite{ond2}.
It is interesting to note that the transition to both of the $2^+$ states can
be interpreted as the knockout of two protons from a $p_{1/2}$ and
a $p_{3/2}$ state.
 Although this has not been directly investigated before, it is natural to 
suppose a connection exists between the spectrum of ${}^{14}{\rm C}$ and the
fragmentation of the $p_{3/2}$ strength in ${}^{15}{\rm N}$. The presence
of a very low-lying $2^+$ state in ${}^{16}{\rm O}$ can also play a role.
The fragmentation of the $p_{3/2}$ strength, in turn, can be interpreted as a
$p_{3/2}$ hole on either the ground state of ${}^{16}{\rm O}$ or one of its
low-lying positive parity states.
 Since these spectra and transition amplitudes are naturally linked to each
other, a formalism in which all of them are obtained in a self-consistent way
is desirable, if not necessary, to resolve the above issues.
Such self-consistent calculations have sofar been restricted to second-order
contributions to the self-energy using a Skyrme force for the effective 
interaction~\cite{vanneck}. Such a treatment of LRC in ${}^{16}{\rm O}$ is 
inadequate since it does not include any residual interaction between the
intermediate 2p1h and 2h1p states in the self-energy.

The merit of the calculations of Refs.~\cite{GeurtsO16,muetherO16} was
that the interaction between the 2p1h states (and 2h1p states) 
was summed to all orders.
Thus a simultaneous description of the effects
of both  particle-hole~(ph) and hole-hole~(hh) (as well as
particle-particle~(pp)) collective excitations was
achieved, including the interplay between them.
  These collective excitations, though, were accounted for only at the
TDA level.  The simultaneous treatment of ph and
pp(hh) excitations is not a trivial problem and an extension of these
calculations beyond TDA presents serious difficulties~\cite{rijs96}. 
 On the other hand, in order to account for the coupling to collective
excitations that are actually observed in ${}^{16}{\rm O}$ it is necessary
to at least consider a random phase approximation (RPA) description of
the isoscalar negative parity states~\cite{czer}.
To account for the low-lying isoscalar positive parity states an even more
complicated treatment will be required~\cite{dickph}.
Sizable collective effects are also present in the particle-particle (pp)
and hh excitations involving tensor correlations for isoscalar and
pair correlations for isovector states. Also in this case, an RPA
treatment would be relevant. 

The aim of the present paper is to extend the calculations of
Ref.~\cite{GeurtsO16} to include such RPA correlations.
A formalism has been proposed recently that allows
the simultaneous inclusion of ph and pp(hh) excitations at the RPA
level~\cite{FaddRPA1,FaddRPA2}.
 In this scheme, the RPA collective
excitations can be summed to all orders by solving a set of Faddeev 
equations for the motion of 2p1h and 2h1p excitations.
 This is done within the framework of self-consistent Green's function
theory (SCGF) in which the equations of motion are expressed in terms of
the dressed (fragmented) sp propagator.
The resulting self-energy contains the inclusion of these pp(hh) and ph
RPA phonons to all orders and therefore allows for an improved description of
the influence of LRC on the sp propagator. In turn, the results
for the hole (and particle) spectral functions can be employed in a 
subsequent dressed RPA (DRPA)
calculation and then iterated to investigate the effects of self-consistency
on the fragmentation.

 In Sec.~\ref{sec:2} we describe the details of the formalism and 
its implementation for ${}^{16}{\rm O}$. The main points of
the Faddeev formalism are reviewed in Sec.~\ref{sec:2fadd}.
 The results for the one-hole spectral function are given in Sec.~\ref{sec:3}.
First, the Faddeev equations are solved including
RPA phonons and a discussion of the improvements over a calculation at the
TDA level is presented.
 Then, in Sec.~\ref{sec:3itr} the RPA results are iterated a few times
to investigate the effects introduced by self-consistent fragmentation.
 In Sec.~\ref{sec:3ph}, the Faddeev formalism is used
to further investigate the relations between the ph spectra
and the sp strength with regard to some of the unresolved questions.
This analysis may give further insight into the understanding of the low-lying
spectra as well as hints for future calculations.
Conclusions are drawn in Sec.~\ref{sec:concl}.

\section{Calculation of the single-particle Green's function}
\label{sec:2}

\begin{figure}
 \begin{center}
 \epsfig{file=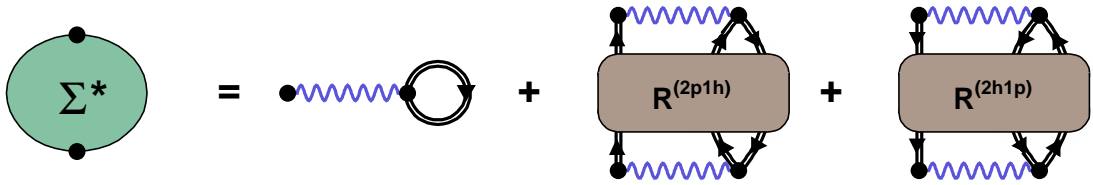, width=16truecm}
 \end{center}
\caption[]{Diagrams contributing to the irreducible self-energy $\Sigma^*$.
   The double lines may represent either an IPM or a dressed propagator. 
The wavy lines correspond to G-matrix interactions.  The first term is the 
   Brueckner-Hartree-Fock potential while the others represent the 2p1h/2h1p
   or higher contributions that are approximated through the Faddeev TDA/RPA
   equations.
\label{fig:selfenergy} }
\end{figure}

We consider the calculation of the sp Green's function
\begin{equation}
 g_{\alpha \beta}(\omega) ~=~ 
 \sum_n  \frac{ \left( {\cal X}^{n}_{\alpha} \right)^* \;{\cal X}^{n}_{\beta} }
                       {\omega - \varepsilon^{+}_n + i \eta }  ~+~
 \sum_k \frac{ {\cal Y}^{k}_{\alpha} \; \left( {\cal Y}^{k}_{\beta} \right)^*  }
                       {\omega - \varepsilon^{-}_k - i \eta } \; ,
\label{eq:g1}
\end{equation}
from which both the one-hole and one-particle spectral functions, for the
removal and addition of a nucleon, can be extracted.
In Eq.~(\ref{eq:g1}),
${\cal X}^{n}_{\alpha} = {\mbox{$\langle {\Psi^{A+1}_n} \vert $}}
 c^{\dag}_\alpha {\mbox{$\vert {\Psi^A_0} \rangle$}}$%
~(${\cal Y}^{k}_{\alpha} = {\mbox{$\langle {\Psi^{A-1}_k} \vert $}}
 c_\alpha {\mbox{$\vert {\Psi^A_0} \rangle$}}$) are the
spectroscopic amplitudes for the excited states of a system with
$A+1$~($A-1$) particles and the poles $\varepsilon^{+}_n = E^{A+1}_n - E^A_0$%
~($\varepsilon^{-}_k = E^A_0 - E^{A-1}_k$) correspond to the excitation
energies with respect to the $A$-body ground state.
The indices $n$ and $k$ enumerate the fragments associated with the
one-particle and  one-hole excitations, respectively.
The one-body Green's function can be
computed by solving the Dyson equation
\begin{equation}
 g_{\alpha \beta}(\omega) =  g^{0}_{\alpha \beta}(\omega) \; +  \;
   \sum_{\gamma \delta}  g^{0}_{\alpha \gamma}(\omega) 
     \Sigma^*_{\gamma \delta}(\omega)   g_{\delta \beta}(\omega) \; \; ,
\label{eq:Dys}
\end{equation}
where the irreducible self-energy $\Sigma^*_{\gamma \delta}(\omega)$ acts
as an effective, energy-dependent, potential. The latter can be expanded in a
Feynman-Dyson series~\cite{fetwa,AAA} in terms of either the mean-field IPM
propagator $g^{0}_{\alpha , \beta}(\omega)$ or even the exact
propagator $g_{\alpha \beta}(\omega)$, which itself is a solution
of~(\ref{eq:Dys}).
 In general, $\Sigma^*_{\gamma  \delta}(\omega)$ can be represented as the sum
of a one-body Hartree-Fock potential and terms that describe the coupling
between the sp motion and more complex excitations~\cite{Win72}.
 This separation is depicted by the diagrams of Fig.~\ref{fig:selfenergy}.
In particular, the latter contributions can be expressed in terms of
a three-line irreducible propagator $R(\omega)$ which describes
the propagation of at least 2h1p or 2p1h at the same time.
 It is at the level of $R(\omega)$ that the correlations involving
interactions between different collective modes need to be included.

 The spectroscopic factors $Z_k$ for the removal
of a nucleon from the $A$ particle system, while leaving
the residual nucleus in its $k$-th excited state, is obtained from the
 spectroscopic amplitudes ${\cal Y}^k_{\alpha}$. The latter
are normalized by
\begin{equation}
 Z_k = \sum_{\alpha}
\left| {\cal Y}^{k}_{\alpha} \right|^2
    =  1 + 
     \sum_{\alpha , \beta}   \left( {\cal Y}^{k}_{\alpha} \right)^* 
       \left.  \frac{ \partial \Sigma^*_{\alpha \beta}(\omega) }
                    { \partial \omega}
       \right|_{\omega = \varepsilon^{-}_k }
	       {\cal Y}^{k}_{\beta}  \; \; .
\label{eq:norm}
\end{equation}
This result follows directly from the Dyson equation~(\ref{eq:Dys}).
 The same relation applies also to the one-particle spectroscopic
amplitudes~${\cal X}^{n}_{\alpha}$.
 The present calculations were performed within a finite set of harmonic
oscillator states,  representing the closed shells that are most relevant
for low-lying excitations.
 As a consequence of the truncation of the
model space a Brueckner G-matrix was employed in evaluating the diagrams
of Fig.~\ref{fig:selfenergy}. The calculation of the Hartree-Fock term and
the approximation of the 2p1h/2h1p propagator are discussed in the following.

\subsection{Brueckner-Hartree-Fock self-energy}

 The first diagram on the right-hand side of 
Fig.~\ref{fig:selfenergy} represents the Brueckner-Hartree-Fock (BHF)
contribution to the self-energy. This is given analytically by
\begin{equation}
 \Sigma^{BHF}_{\alpha \beta}(\omega) =
   i \sum_{\gamma \delta} \int \frac{d \omega '}{2 \pi} 
   G_{\alpha \gamma, \delta \beta}(\omega + \omega ')
      g_{\gamma \delta}(\omega ') ,
\label{eq:BHF}
\end{equation}
where the energy dependence is a consequence of the use of the Brueckner 
G-matrix.
The Pauli operator used in the calculation of the G-matrix excludes
all the intermediate states that are part of the model
space, in which the LRC are explicitly computed~\cite{muetherO16}.
This double-partitioning procedure avoids double counting of the pp 
ladder-diagram contributions to the self-energy.
The self-energy contribution of Eq.~(\ref{eq:BHF})
is needed in order to generate the correct sp energies for the main
shells. Nevertheless, it should be noted that Eq.~(\ref{eq:BHF}) is expressed
in terms of the self-consistent solution~$g_{\gamma \delta}(\omega)$.
If instead an IPM input is used, the approximated Hartree-Fock contribution
may not be sufficient to put the main hole and particle
fragments at the right place in energy.
 Rather, a self-consistent solution of the BHF equations should be employed
to evaluate Eq.~(\ref{eq:BHF}). 
 For example, in Ref.~\cite{GeurtsO16}, this is done by solving
the BHF equations in advance. Then, the set of precomputed sp
energies was used in the rest of the calculations.

  In the present work, this issue was solved by adding to the
self-energy auxiliary one-body potential, diagonal in the model space. This
makes up for the deficiencies introduced by the lack of self-consistency in
Eq.~(\ref{eq:BHF}). 
 Only the few matrix elements relative to fragments close to the Fermi energy
needed to be set different from zero when the IPM starting point was used.
 These parameters were fitted to reproduce the correct missing energies for 
the knockout and the addition of a proton.
  When the calculations are iterated, the term~(\ref{eq:BHF})
produces the correct BHF potential and no need for further corrections is
needed. Accordingly, in our calculation the corrections applied 
to the $s$ and $d$ shells became negligibly small after a few iterations.
 The $p$ shells continue to require an adjustment 
of 2.7~MeV for $1p_{1/2}$ and -0.7~MeV for
$1p_{3/2}$, respectively, mainly to obtain the correct spin-orbit splitting.

 The BHF contribution~(\ref{eq:BHF}) is
also relevant for the normalization of the spectral strengths, since
it contributes to the derivative of the self-energy in Eq.~(\ref{eq:norm}).
 In Ref.~\cite{GeurtsO16}, it was shown that this accounts for a proper
treatment of the depletion induced by SRTC at least for the normally
occupied shells in the IPM.
 In the present work, the energy dependence of the BHF contribution
was taken into account both in solving the Dyson equation and
in the normalization of the spectral amplitudes.

\subsection{Faddeev approach to the self-energy}
\label{sec:2fadd}

 In order to compute the last diagrams of Fig.~\ref{fig:selfenergy}, we first
consider the polarization propagator describing excited states in the system 
with $A$ particles
\begin{eqnarray}
 \Pi_{\alpha \beta , \gamma \delta}(\omega) &=& 
 \sum_{n \ne 0}  \frac{  {\mbox{$\langle {\Psi^A_0} \vert $}}
            c^{\dag}_\beta c_\alpha {\mbox{$\vert {\Psi^A_n} \rangle$}} \;
             {\mbox{$\langle {\Psi^A_n} \vert $}}
            c^{\dag}_\gamma c_\delta {\mbox{$\vert {\Psi^A_0} \rangle$}} }
            {\omega - \left( E^A_n - E^A_0 \right) + i \eta } 
\nonumber \\ & & 
~+~ \sum_{n \ne 0} \frac{  {\mbox{$\langle {\Psi^A_0} \vert $}}
              c^{\dag}_\gamma c_\delta {\mbox{$\vert {\Psi^A_n} \rangle$}} \;
                 {\mbox{$\langle {\Psi^A_n} \vert $}}
             c^{\dag}_\beta c_\alpha {\mbox{$\vert {\Psi^A_0} \rangle$}} }
            {\omega - \left( E^A_0 - E^A_n \right) - i \eta } \; ,
\label{eq:Pi}
\end{eqnarray}
and the two-particle propagator relevant for the $A \pm 2$ excitations
\begin{eqnarray}
 g^{II}_{\alpha \beta , \gamma \delta}(\omega) &=& 
 \sum_n  \frac{  {\mbox{$\langle {\Psi^A_0} \vert $}}
                c_\beta c_\alpha {\mbox{$\vert {\Psi^{A+2}_n} \rangle$}} \;
                 {\mbox{$\langle {\Psi^{A+2}_n} \vert $}}
         c^{\dag}_\gamma c^{\dag}_\delta {\mbox{$\vert {\Psi^A_0} \rangle$}} }
            {\omega - \left( E^{A+2}_n - E^A_0 \right) + i \eta }
\nonumber \\  
&+& \sum_k  \frac{  {\mbox{$\langle {\Psi^A_0} \vert $}}
    c^{\dag}_\gamma c^{\dag}_\delta {\mbox{$\vert {\Psi^{A-2}_k} \rangle$}} \;
                 {\mbox{$\langle {\Psi^{A-2}_k} \vert $}}
                  c_\beta c_\alpha {\mbox{$\vert {\Psi^A_0} \rangle$}} }
            {\omega - \left( E^A_0 - E^{A-2}_k \right) - i \eta } \; .
\label{eq:g2}
\end{eqnarray}
In their Lehmann representations, these quantities contains all the
relevant information regarding ph and pp(hh) collective excitations.
 The approach of Ref.~\cite{FaddRPA1} consists in computing these quantities
by solving the ph-TDA/RPA and the pp-TDA/RPA equations~\cite{schuckbook},
respectively. 
 In the most general case of a fragmented input propagator, the
corresponding dressed RPA/TDA (DRPA/DTDA) equations~\cite{DRPAg,DRPApaper}
are solved.
 After obtaining the propagators~(\ref{eq:Pi}) and~(\ref{eq:g2}),
they have to be coupled to the sp motion in order to obtain 
the corresponding approximation 
for the forward- and backward-going $R(\omega)$ propagators.
 This is achieved by solving two separate sets of Faddeev equations for
the 2p1h and the 2h1p propagation, respectively.

\begin{figure}
 \begin{center}
 \epsfig{file=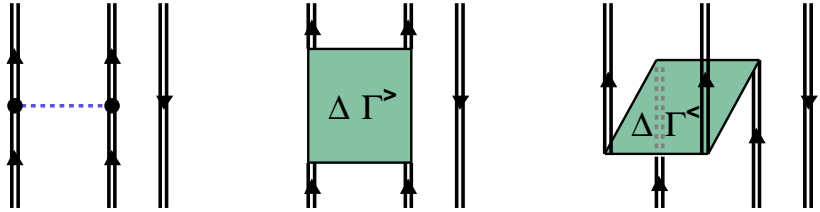, width=12truecm}
 \end{center}
\caption[]{Faddeev vertex for the coupling of a pp excitation to a hole line.
         Here $\Delta \Gamma^>$ and $\Delta \Gamma^<$ represent the forward-
	and backward-going part of the pp DRPA propagator.}
\label{fig:Gpp} 
\end{figure}

 Taking the 2p1h case as an example, one can define three different
components $R^{(i)}(\omega)$ ($i=1,2,3$) that differ from each other
by the last interaction that appears in their diagrammatic expansion. These
components are solutions of the Faddeev equations~\cite{Fadd1,gloeck}
\begin{eqnarray}
  \lefteqn{
  R^{(i)}_{\mu     \nu    \lambda  ,
           \alpha  \beta  \gamma   }(\omega) }
        \hspace{.3in} & &
\nonumber  \\
   &=&  \frac{1}{2}\left( 
      {G^0}^>_{\mu      \nu     \lambda   ,
               \alpha   \beta   \gamma   }(\omega)
    - {G^0}^>_{\nu      \mu     \lambda   ,
               \alpha   \beta   \gamma   }(\omega) \right)
\nonumber  \\
  & + & ~ {G^0}^>_{\nu   \mu     \lambda  ,
                  \mu'  \nu'    \lambda' }(\omega) ~
   \Gamma^{(i)}_{\nu'   \mu'     \lambda'  ,
                 \mu''  \nu''    \lambda'' }(\omega) ~
\nonumber  \\
  & \times & ~
  \left( R^{(j)}_{\mu''   \nu''  \lambda''  ,
                  \alpha  \beta  \gamma    }(\omega) ~+~
         R^{(k)}_{\mu''   \nu''  \lambda''  ,
                  \alpha  \beta  \gamma    }(\omega)
       \right)  \;  ,
\label{eq:FaddTDA}
\end{eqnarray}
 in which repeated indices are summed over and ($i,j,k$) are cyclic
permutations of ($1,2,3$).
 In Eq.~(\ref{eq:FaddTDA}), ${G^0}^>(\omega)$ is the forward-going part
of the 2p1h propagator for three noninteracting lines.
 Using the notation introduced in Eq.~(\ref{eq:g1}), we have
\begin{equation}
  {G^0}^>_{\mu     \nu    \lambda  , 
           \alpha  \beta  \gamma  }(\omega) ~=~
        \sum_{n_1 , n_2 , k} ~
     \frac{ \left( 
         {\cal X}^{n_1}_{\mu}
         {\cal X}^{n_2}_{\nu}
         {\cal Y}^{k}_{\lambda}
            \right)^* \; 
         {\cal X}^{n_1}_{\alpha}
         {\cal X}^{n_2}_{\beta}
         {\cal Y}^{k}_{\gamma} }
    { \omega - ( \varepsilon^+_{n_1} + \varepsilon^+_{n_2} -
                                     \varepsilon^-_{k} ) + i \eta } \; .
\label{eq:G0fw}
\end{equation}
The Faddeev vertices
$\Gamma^{(i)}(\omega)$  contain the couplings of a ph or pp(hh) collective
excitation and a freely propagating line. 
 A well known characteristic of the Faddeev formalism is that it introduces
spurious solutions in addition to the correct eigenstates of the Schr\"odinger
equation~\cite{gloeck99}.  In Ref.~\cite{FaddRPA1} it was shown that
this issue imposes constraints on the choice of the
vertex $\Gamma^{(i)}(\omega)$ terms and as a consequence both the forward-
and backward-going parts of the collective RPA excitations
should be included in these terms. Fig.~\ref{fig:Gpp} shows an example of the
diagrammatic expansion of the vertex corresponding to attaching a
hole line to a pp(hh) phonon. 
 By summing over all the Faddeev components (and subtracting
${G^0}^>(\omega)$ to avoid double counting) one finally obtains the
2p1h propagator
 \begin{eqnarray}
    R^{(2p1h)}_{\mu \nu \lambda , \alpha \beta \gamma}(\omega) &=&
   \sum_{i=1,2,3}     R^{(i)}_{\mu \nu \lambda , \alpha \beta \gamma}(\omega)
 \nonumber \\
 &-& \frac{1}{2} \left( {G^0}^>_{\mu \nu \lambda , \alpha \beta \gamma}(\omega)
        - {G^0}^>_{\nu \mu \lambda , \alpha \beta \gamma}(\omega) \right) \; ,
\label{eq:faddfullR}
\end{eqnarray}
from which the self-energy can be easily derived.
Al the above consideration also apply to the 2h1p propagation, for which a set
of equations analogous to~(\ref{eq:FaddTDA}), (\ref{eq:G0fw})
and~(\ref{eq:faddfullR}) is employed.

\begin{figure}
 \begin{center}
 \epsfig{file=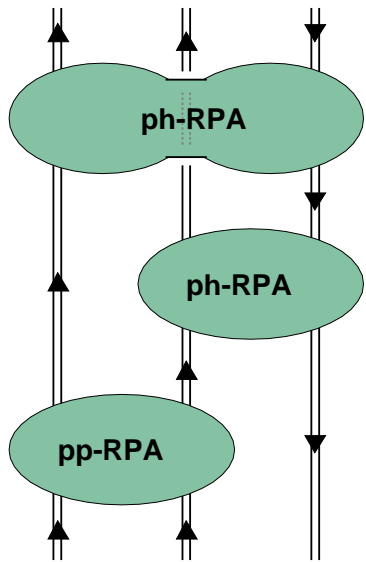, width=6truecm}
 \end{center}
\caption[]{Example of diagrams that are summed to all orders by means of 
the Faddeev equations.}
\label{fig:FaddSum} 
\end{figure}

 The actual application of the Faddeev formalism to 2p1h/2h1p propagation
involves a certain number of  complications.
  These rewquire a slight redefinition
of the components $R^{(i)}(\omega)$ introduced above and a rearrangement
of Eqs.~(\ref{eq:FaddTDA}). Eventually one is left which an eigenvalue problem
that can be projected on a Hilbert space that spans only the correct physical
solutions.
 The details of these issues have already been presented in 
Ref.~\cite{FaddRPA1} and will not be discussed any further in this paper.
 The important thing to note here is that this formalism allows the inclusion
of the effects of ph and pp(hh) motion not only at the TDA level but also
at the more collective RPA level.
 These excitations are coupled to each other by the
Faddeev equations, generating diagrams like the one shown in
Fig.~\ref{fig:FaddSum}. This also assures that Pauli correlations are properly
taken into account at the 2p1h/2h1p level.
In addition, one can employ dressed sp propagators in these
equations to generate a self-consistent solution.

\subsection{Application to ${}^{16}{\rm O}$ and iterative procedure}
\label{sec:2itr}

 In the calculations described below, the Dyson equation was solved within
a model space consisting of harmonic oscillator sp states.
An oscillator parameter $b=1.76 fm$ was chosen and all the first four major
shells (from $1s$ to $2p1f$) plus the $1g_{9/2}$ where included.
 Inside this model space, the interaction used was a Brueckner
G-matrix~\cite{CALGM} derived from the Bonn-C potential~\cite{bonnc}.
 This version of the Bonn potential does not include
any charge independence breaking term and the Coulomb interaction between
protons was not taken into account.
 Therfore the same results were obtained for both neutron and proton spectral
functions.
 As mentioned above the energy dependence of the G-matrix was completely
taken into account in the evaluation of the BHF part of the self-energy and
in computing the normalization of spectral amplitudes.
For the solution of the Faddeev equations a G-matrix evaluated at a
fixed starting energy was employed. 
  In this case a value of -25~MeV has been chosen as a suitable
average of the most important 2h1p states which are of interest here.

By using an IPM propagator as input Green's function,
the BHF term~(\ref{eq:BHF}) was computed and the Faddeev equations were solved
to obtain the irreducible self-energy of Fig.~\ref{fig:selfenergy}.
From the solution of Eq.~(\ref{eq:Dys}) the sp
spectral functions were obtained.
The solution to the Dyson equation~(\ref{eq:Dys}) contains a large number of
fragments, most of which are quite small.
 A fully self-consistent solution requires a method
in which the sp strength is binned over a large energy domain~\cite{vanneck}.
The number of poles and the resulting 2p1h and 2h1p are then too numerous
for a practical solution of the Faddeev equations.
To obtain some insight into the effects of self-consistency we have chosen the
following procedure to account for the fragmentation.
For sp levels far from the Fermi energy, we have kept two poles above and
below the Fermi energy, except for the $f$ and $g$ shells for which only one 
effective hole pole was kept. For the levels near the
Fermi energy, the quasiparticle (hole) fragment was kept, including its 
location and strength.
When two effective poles on one side of the Fermi energy were included,
the fragment closest
to the Fermi energy was kept with its strength and the rest of the strength
was collected at a location determined by weighing the remaining fragment
energies with the corresponding strength.
The resulting dressed propagator still contains all the relevant low-energy
fragmentation and at the same time it has a number of poles small enough to be
used as input in another Faddeev DRPA calculation.

In performing the iterations, the ph DRPA equations that give the spectrum
of ${}^{16}{\rm O}$, generate an instability in the isoscalar $0^+$ channel.
The RPA equations using IPM input do not generate this instability.
No other instabilities are encountered in other channels, including pp or hh
ones.
Naturally, the instability of the lowest ph $0^+$ state tends to disappear
when a 
more negative starting energy is chosen for the G-matrix since such a choice 
reduces the attraction in this channel.
  Since the first $0^+$ state is of particular importance,  we decided not
to discard it but to compute it in a regime 
were the instability disappears. A stable solution for the spectroscopic
amplitudes of this state was obtained by solving the ph DPRA equations
with a G-matrix at a starting energy of -110~MeV. 
The energy of the state was then kept fixed
at the experimental energy of 6.05~MeV.
  All the remaining $0^+$ levels were properly computed with a G-matrix
at -25~MeV.

\section{Results for the one-hole spectral function}
\label{sec:3}

\subsection{Effects of RPA correlations}

 By using an IPM ansatz as input propagator, the Faddeev equations have been
applied to obtain the self-energy of ${}^{16}{\rm O}$ in both TDA and RPA
approximations. 
 The resulting self-energy was then used in the
Dyson equation~(\ref{eq:Dys}) to obtain the sp
spectral functions for the removal (one-hole) and addition
(one-particle) of a nucleon.
 The values of the spectroscopic factors for the main particle and hole shells
close to the Fermi energy are reported in Table~\ref{tab:IPMpeaks}.
  The hole strengths given by TDA are 0.775 for $p_{1/2}$ and 0.766
for $p_{3/2}$, in close agreement with the results of Ref.~\cite{GeurtsO16} 
(to which the present TDA calculation is equivalent).
 The introduction of RPA correlations reduces these values and brings them
down to 0.745 and 0.725, respectively.
This result reduces the discrepancy with the experiment by about 4\% and
shows that collectivity beyond the TDA level is relevant to
explain the quenching of spectroscopic factors.
It should be noted that the present RPA results describing the
ph and pp (hh) spectrum suffer from the usual problems associated with
RPA.
One such feature, as already noted above for the ph $0^+$ state, is the 
appearance of at most one collective state for a given $J^{\pi}$, whereas
many low-lying isoscalar natural parity states are observed experimentally. 
This feature implies that especially the ph spectrum does not provide
a very good description of the experimental data.
One may therefore hope that further improvements of the description
of the RPA phonons themselves will close the gap with the experimental
data further.

 Together with the main fragments, the Dyson equation produces also a
large number of solutions with small spectroscopic factors.
 For the one-hole spectral function, this strength
extends down to about -130~MeV. This background partly represents 
the strength that is removed from the main peaks
and shifted up to medium missing energies.
The energy dependence of the G-matrix accounts for another 10\% effect
in pushing the strength of the mostly occupied shells to high energies
in the particle domain. We note that the location of this
strength cannot be expicitly calculated in the present approach but
corresponds to very large energies~\cite{bv91}.
 The occupation number coming from both the background contribution
at negative energies, as well as the main
hole fragments is displayed in 
Table~\ref{tab:IPMstrengths} for the most important shells.
 Summing these numbers together with the occupation of the main peaks
and weighing them by a factor of $2(2j+1)$, one gets a total number
of particles equal to about 15 nucleons.  This violation of particle number
is a consequence of the energy dependence of the G-matrix.
The remaining strength is then accounted for by the presence of high-momentum
components due to SRTC not explicitly calculated in the present scheme.
The present result therefore also gives 
an estimate of the number of these high-momentum particles that are
shifted to even higher energies (more negative).
These high-momentum components are included in the results of 
Refs.~\cite{md94,pmd95} and their strength corresponds to the number
of missing nucleons in the present calculation.
The effects of SRTC on the reduction of quasihole spectroscopic factors are 
properly included, through the energy dependence of the G-matrix
interaction.

\begin{figure}
 \begin{center}
 \epsfig{file=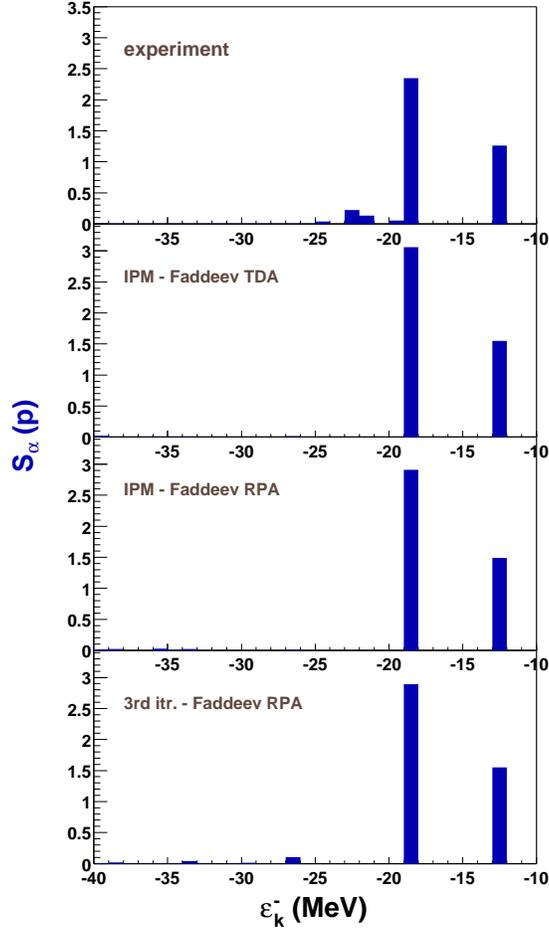, width=8truecm}
 \end{center}
\caption[]{One-proton removal strength as a function of the 
   hole sp energy $\varepsilon^{-}_k = E^A_0 - E^{A-1}_k$
    for ${}^{16}{\rm O}$ and angular momentum
  $\ell =1$.  The experimental values are taken from~\cite{leus}. The theoretical
   results
   have been calculated in both TDA and RPA approximation with an IPM model
   input.  The bottom panel includes the results of iterating the
fragmentation pattern through the construction of DRPA phonons.
\label{fig:phsf} }
\end{figure}

\begin{figure}
 \begin{center}
 \epsfig{file=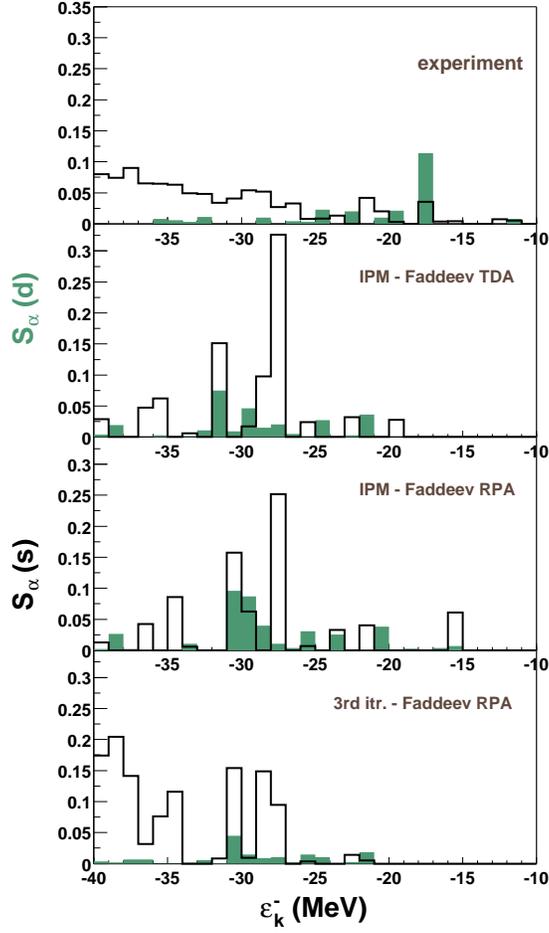, width=8truecm}
 \end{center}
\caption[]{One-proton removal strength as a function of the 
   hole sp energy $\varepsilon^{-}_k = E^A_0 - E^{A-1}_k$
    for ${}^{16}{\rm O}$ and positive parity
   final states. The solid bars correspond
to results for orbital angular momentum $\ell =2$, while the
   thick lines refer to $\ell =0$.
    The experimental values are taken from~\cite{leus}. The theoretical
    results
   have been calculated in both TDA and RPA approximation with an IPM model
   input.  The bottom panel includes the effect of fragmentation on the
construction of the DRPA phonons after three iterations.
\label{fig:sdhsf} }
\end{figure}

Fig.~\ref{fig:phsf} displays the TDA and RPA one-hole spectral function
for the $p_{1/2}$ and $p_{3/2}$ states.
In this figure the theoretical spectral
function is binned in order to make a comparison with the experimental results.
These results demonstrate that neither of the two approaches explains the
breaking of the main $p_{3/2}$ peak when an input IPM propagator is used.
 The main difference between the two results are the 4\% smaller peaks
obtained in the RPA approach.
 The results for positive parity shells are shown in Fig.~\ref{fig:sdhsf}.
The solid bars refer to results for orbital angular momentum $\ell = 2$
and the open bars to $\ell =0$, respectively.
We observe that the RPA approach generates two 
hole peaks with
angular momenta $d_{5/2}$ and $s_{1/2}$ at small missing energy.
 These peaks are
found separated from the rest of strength at  -15.6 and -15.8~MeV
respectively, which differ from the experimental value by about 2~MeV. 
 The theory also  predicts a spectroscopic factor of 0.1\% for $d_{5/2}$
which is smaller than the experimental value of 1.9\%.
 This represents an improvement with respect
to the TDA, where such a fragment is not reproduced at all.
 The agreement is better with the $s_{1/2}$ fragment for which the theory
predicts 3.0\% and the experimental value is 1.8\%.
 At energies below -20~MeV, the experimental $s_{1/2}$ strength is
distributed almost continuously and increases as the energy approaches
the region corresponding to giant resonances.
 In the present calculation, based on a finite number of discrete states,
the theory predicts a fragmentation over fewer isolated peaks with
higher spectroscopic strengths.

\subsection{Effects of fragmentation}
\label{sec:3itr}

The RPA results were iterated a few times, with the aim of studying
the effects of fragmentation on the RPA phonons and, subsequently,
on the spectral strength.
 This was done by employing the prescription for representing the strength
distribution with effective poles that is described in Sec.~\ref{sec:2itr}.
The negative parity hole spectral function resulting from the third
iteration is shown in the lower panel of Fig.~\ref{fig:phsf}. 
 The main difference between these results and the one obtained by using
an IPM input is the appearance of a second smaller $p_{3/2}$ fragment at 
-26.3~MeV. This peak rises in the first two iterations and appears
to become stable in the last one, with a spectroscopic factor of 2.6\%.
 This can be interpreted as a peak that describes the fragments seen
experimentally at slightly lower energy.
  This is the first time that such a fragment is
obtained in calculations of the spectral strength.
Further insight into the appearance of this strength is discussed
in Sec.~\ref{sec:3ph}.

 A second effect of including fragmentation in the construction of the RPA 
phonons is to increase the strength of the main hole
peaks.  The spectroscopic factors for the main $p$ peaks, as obtained from
different iterations, are reported in Table~\ref{tab:123itr}.  The $p_{1/2}$ 
strength increases from the 0.745 obtained with IPM input to 0.774,
essentially cancelling the improvement gained by 
the introduction of RPA correlations over the TDA ones.
 The main peak of the $p_{3/2}$ remains at 0.722 but the appearance of the
secondary fragment slightly increases the overall strength at low
energy also.
 This behavior can be understood by realizing that with an IPM input
most of the phonons are somewhat more collective than the ones obtained
from employing dressed propagators with the exception of the special case of
the ph $0^+$.
As a result, one can expect a reduced effect of RPA correlations when
fragmentation is included in the construction of the phonons.
This feature has also been observed in other self-consistent calculations
of the sp spectral strength, for example in nuclear matter~\cite{libth}.
 Obviously, this makes the disagreement with experiments a little worse and
additional work is needed to resolve the disagreement with the data.
Nevertheless it is clear 
that fragmentation is a relevant feature of nuclear systems and that
it has to be properly taken into account.
 It is also worth nothing that already after a few iterations, all the main
quantities of Table~\ref{tab:123itr} tend to stabilize and sustain themselves
in a self-consistent way.

 Table~\ref{tab:123itr} also shows the total number of particles obtained at each
iteration (derived by summing over the hole strength).
 This result corresponds to about 14.6 nucleons when fragmentation is 
included.
This gives an estimate for the overall occupancy of high momentum states of
about 10\%, in agreement with direct calculations of SRTC~\cite{md94,pmd95}.
We observe that this estimate is different from the results for the IPM input
quoted in Table~\ref{tab:IPMstrengths}.
We associate this difference with the energy dependence of the G-matrix
which is sampled differently in both cases.
In the IPM calculation the lowest three shells are included at the 
harmonic oscillator level. Upon iteration, which involves the changing
BHF contribution, the admixture of the other $s$ and $p$ shells 
is included and will generate a slightly different effect related to
the energy dependence of the G-matrix since different matrix elements
are sampled in each case.
We note here that for this reason there is also a distinct difference
between the quasiparticle and quasihole strengths near the Fermi energy
as shown in Table~\ref{tab:IPMpeaks} of about 10\%.
This same difference appears in the summed strengths appearing in 
Table~\ref{tab:IPMstrengths}.
In both cases there appears more strength in the particle domain that is
appropriate for the effect of SRTC.
In the present approach we cannot treat this effect for particle shells
properly since the G-matrix can only be calculated reliably at negative energy.
The derivative of the energy dependence of the G-matrix at energies
relevant for particle states will therefore not reflect the true
depletion due to SRTC.
For this reason the summed strength for the particle states is close to
1 in Table~\ref{tab:IPMstrengths}.

 The results of the third iteration are also given in
Fig.~\ref{fig:sdhsf} for the relevant positive parity spectral functions.
We note that the $s_{1/2}$ and $d_{5/2}$ hole fragments at -15~MeV are
no longer generated by these iterated calculations.
Also, as a consequence of dressing the input propagator, more poles are
produced as solutions of the Dyson equation~(\ref{eq:Dys}). This allows for 
a better distribution of the $s_{1/2}$ strength at medium missing energies.
Similar results have been obtained in the self-consistent second-order
calculation for ${}^{48}{\rm Ca}$ in Ref.~\cite{vanneck}.
Presumably, a more complete representation of the strength in the input
propagator would further improve the $\ell =0$ strength distribution.

\subsection{Role of $0^+$ and $3^-$ excited states in ${}^{16}{\rm O}$}
\label{sec:3ph}

\begin{figure}
 \begin{center}
 \epsfig{file=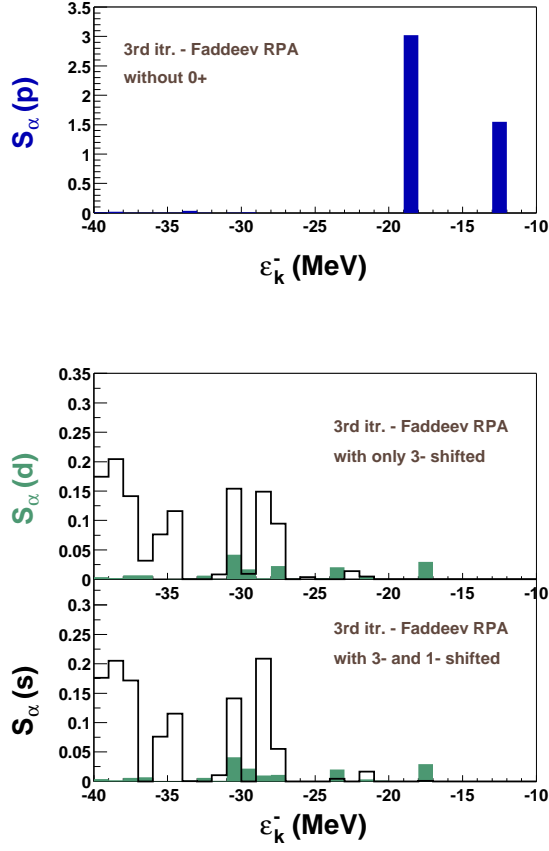, width=8truecm}
 \end{center}
\caption[]{One-proton removal strength resulting from repeating the third
     iteration with a modified ph propagator.  The upper panel refers to
  the results for $\ell =1$ when the lowest $0^+$ state of ${}^{16}{\rm O}$
     is excluded.
  The lower panels give the $\ell =0$ (thick curve) and $\ell =2$ (solid bar)
     results
     obtained when the $3^-$ alone or both the $3^-$ and $1^-$ states are
     shifted to their experimental energies.
\label{fig:no0+} }
\end{figure}

 A deeper insight into the mechanisms that generate the fragmentation
pattern can be
gained by looking directly at the connection between the spectral function
and some specific collective states.
 To clarify this point we repeated the third iteration using exactly the same
input but without replacing the unstable ph $0^+$ state, which was instead
discarded.
 The resulting $p$ hole spectral function is shown in the upper panel of
Fig.~\ref{fig:no0+}.
 In this calculation no breaking of the $p_{3/2}$ peak is obtained. Instead 
a single peak is found with a spectroscopic factor equal to 0.75 which
corresponds to the sum of the two fragments that are obtained when the 
$0^+$ state is taken into account.
 This result can be interpreted by considering the $p_{3/2}$ fragments as a
hole on the ground state and on the first excited $0^+$ state of the
${}^{16}{\rm O}$ core, respectively.
 If the latter two levels are close enough to each other in the calculation,
a mixing between the two configurations can occur and a second
smaller fragment is generated.
 When the excited $0^+$ state is removed from the calculation or, like
in the TDA approach, is found far above the experimental energy, the
calculation can reproduce only one single peak.
 Obviously, it is understood from the dressed results of Fig.~\ref{fig:phsf}
that further improvements have to be made in other to describe properly the
strength and the missing energy of the smaller fragments.
 A candidate to consider in this improvement is the role of
the first $2^+$ state in ${}^{16}{\rm O}$, which can also couple
to generate $p_{3/2}$ hole fragments but that was not included here since
it cannot be obtained by the present ph DRPA calculation, at least not
at low enough energy.

The other two low-lying states of ${}^{16}{\rm O}$ that may be of some
relevance are the isoscalar $1^-$ and $3^-$. 
These excitation are reproduced reasonably well
by RPA type calculations~\cite{czer} but are typically found at higher
energies than the experimental ones. In the present case the third iteration
gives 9.4 and 10.8~MeV for $3^-$ and $1^-$ respectively, which is
about 3~MeV above the experimental results.
This points to a need for a more attractive G-matrix interaction.
We mention here
that the present G-matrix is calculated without any binding correction of
the sp energies for particle states which could have some influence on
the strength of the effective interaction.
 The lower panels of Fig.~\ref{fig:no0+} show the results for the 
even parity spectral functions that are obtained if the $3^-$ alone or both
$3^-$ and $1^-$ are shifted down to their experimental values. 
In this case, a $d_{5/2}$ hole peak is obtained at low missing energy. This
result is also quantitatively more satisfactory that what was obtained 
in the RPA calculation based on the IPM, since in this case it is found
at -17.7~MeV (in agreement with 
experiment) and with a spectroscopic strength of 0.5\%.
 It is interesting to note that the shifting of the $3^-$ collective state
does not produce any other noticeable change in the theoretical spectral
function.  The same applies if also the $1^-$ is shifted.

It appears therefore that the main impediment for further improvements 
of the description of the experimental data is associated with the deficiencies
of the RPA (DRPA) description of the excited states. One important problem
is the appearance of at most one collective phonon for a given $J^{\pi},T$
combination. Experimentally, several low-lying isoscalar $0^+$ and $2^+$
excited states are observed at low energy in ${}^{16}{\rm O}$ as well
as additional $3^-$ and $1^-$ states.
A possible way to proceed would be to first concentrate on an improved 
description of the collective phonons by extending the RPA to explicitly
include the coupling to two-particle$-$two-hole (2p2h) states.
Such an extended RPA procedure has been applied in heavier nuclei with
considerable success~\cite{brand88,brand90}.
In order to be relevant for ${}^{16}{\rm O}$, this approach requires
an extension in which the coherence of the 2p2h states is included
in the form of the presence of two phonon excitations.
Such contributions arise naturally when the response is calculated
by using the Baym-Kadanoff construction of the irreducible ph interaction
which is based on a self-consistent treatment of the self-energy~\cite{bk61}.
 
\section{Conclusions}
\label{sec:concl}

 In the present paper the Faddeev technique has been applied to study 
2h1p correlations 
at small missing energies for the nucleus of ${}^{16}{\rm O}$.
The application of the Faddeev method allows for the first time the treatment
of the coupling of ph and hh collective modes 
within an RPA framework and to all orders in the nucleon self-energy.
 The resulting spectral function shows better agreement with experimental
data than all previous calculations. 
Additional encouraging results are obtained in the form low-lying positive
parity fragments.

These results were extended by recalculating the RPA phonons using the
so-obtained fragmented sp propagator.
The inclusion of this fragmentation in the phonons leads to
the appearance of an additional $p_{3/2}$ fragment at low energy
in agreement with experiment.
Other features,
like the presence of positive parity $5/2$
and $1/2$ holes at -17~MeV, cannot be obtained  when the present calculations
are iterated.

We have further identified the important role played by the low-lying ph
states of ${}^{16}{\rm O}$.
The low-lying $0^+$ appears to be at least partially
responsible for the splitting of the $p_{3/2}$ strength at low energy,
whereas the low-lying $3^-$ state plays a decisive role in the presence
of $d_{5/2}$ strength at low energy.
The results of the present calculations are therefore very sensitive
to the quality of the RPA (DRPA) description of the ph spectrum.
It is well known that this description is as yet unsatisfactory but key
ingredients for further improvements can be identified through
the Baym-Kadanoff procedure based on self-consistent propagators.
We therefore conclude that the present 
results show that further improvement in the understanding of
the low-energy fragmentation can be gained. To do this, the employed
approximations need to correctly reproduce all the lowest
ph collective modes. 
 We therefore propose to first improve the quality of the RPA phonons
before engaging in a fully self-consistent evaluation of the one- and
two-hole spectral functions for ${}^{16}{\rm O}$~\cite{inprep}.

\acknowledgments
 Part of the calculations described in this paper were performed on the
SGI Origin 2000 of the Center for Scientific Parallel Computing of
Washington University, St.~Louis.
 One of us (C.B.) would like to acknowledge the hospitality of the
Institut f\"ur Theoretische Physik at the University of T\"ubingen,
where this work has been completed.
This work is supported by the U.S. National Science Foundation
under Grant No.~PHY-9900713.



\begin{table}[t]
    \parbox[t]{1.\linewidth}{
      \caption[]{\label{tab:IPMpeaks}
        Spectroscopic factors for ${}^{16}{\rm O}$ as computed
       in both TDA and RPA schemes using an IPM input.
        Listed are the strengths of the main (particle or hole) fragments
       for the five levels close to the Fermi energy.
        All values are given as a fraction of the corresponding
       IPM value. }
      }
 \begin{tabular}{ccc}
     Shell  &  TDA  &   RPA \\
  \hline 
 $d_{3/2}$  & 0.866 &  0.838   \\
 $s_{1/2}$  & 0.882 &  0.842  \\
 $d_{5/2}$  & 0.894 &  0.875   \\
 $p_{1/2}$  & 0.775 &  0.745  \\
 $p_{3/2}$  & 0.766 &  0.725  \\
 \end{tabular}
\end{table}

\begin{table}[t]
    \parbox[t]{1.\linewidth}{
      \caption[]{\label{tab:IPMstrengths}
        
  Occupation and depletion numbers for the most
       relevant shells of ${}^{16}{\rm O}$ as computed
       in both TDA and RPA schemes using an IPM input.
         All the results are given as a
       fraction of the corresponding IPM value.
        Also shown is the result for the total number of nucleons
       obtained by summing over all the hole fragments. }
      }
 \begin{tabular}{ccccc}
     & \multicolumn{2}{c}{TDA} & \multicolumn{2}{c}{RPA} \\
    Shell   & Particle &  Hole   & Particle &  Hole \\
  \hline 
 $2p_{1/2}$    & 0.983 & 0.014    &  0.980  & 0.017 \\
 $2p_{3/2}$    & 0.980 & 0.016    &  0.978  & 0.018 \\
 $1d_{3/2}$    & 0.958 & 0.038    &  0.945 &  0.051  \\
 $2s_{1/2}$    & 0.954 & 0.039    &  0.916  & 0.074 \\
 $1d_{5/2}$    & 0.961 & 0.035    &  0.946 &  0.049  \\
 $1p_{1/2}$    & 0.102 & 0.828    &  0.128  & 0.804 \\
 $1p_{3/2}$    & 0.076 & 0.856    &  0.107  & 0.828 \\
 $1s_{1/2}$    & 0.044 & 0.888    &  0.057  & 0.876 \\
  \hline 
 \multicolumn{2}{l}{Total occ.} &  14.95      &     & 15.06 \\
 \end{tabular}
\end{table}

\begin{table}[t]
    \parbox[t]{1.\linewidth}{
      \caption[]{\label{tab:123itr}
  Hole spectroscopic factors ($Z_{\alpha}$) for knock out of a $\ell =1$
      proton from ${}^{16}{\rm O}$ and occupation
      numbers ($n_{\alpha}$) for different angular momenta of the nucleon.
       These results refer to the first three iterations of the DRPA equations.
       All the values are given as a fraction of the corresponding IPM value
      and in the case of $\ell =0$ and $\ell =1$ are summed over the principal
      h.o. quantum numbers belonging to the model space (i.e. $s_{1/2}$ stands
      for the sum of $1s_{1/2}$ and $2s_{1/2}$, similary for $p_{1/2}$
      and $p_{3/2}$). Also included is the total number of nucleons for each 
iteration.}
      }
 \begin{tabular}{lrrr}
    Shell  & 1st itr. & 2nd itr. & 3rd itr. \\
  \hline 
 $Z_{p_{1/2}}$ & 0.775 &  0.777  & 0.774  \\  \\
 $Z_{p_{3/2}}$ & 0.725 &  0.727  & 0.722  \\
               & 0.015 &  0.027  & 0.026  \\ \\
  \hline  \\
 $n_{d_{3/2}}$ & 0.025 &  0.025 &  0.026  \\
 $n_{d_{5/2}}$ & 0.020 &  0.021 &  0.021  \\

 $n_{p_{1/2}}$ & 0.850 &  0.848  & 0.848  \\ 

 $n_{p_{3/2}}$ & 0.870 &  0.871  & 0.870  \\ 

 $n_{s_{1/2}}$ & 0.911 &  0.914  & 0.916  \\
  \hline 
  Total occ.  & 14.56 &  14.57   & 14.58  \\
 \end{tabular}
\end{table}

\end{document}